\documentclass[twocolumn,showpacs,preprintnumbers,amsmath,amssymb,aps,showkeys]{revtex4}
\usepackage{graphics}% Include figure files

\begin{document}
\title{Modified Newtonian Dynamics as an extra dimensional effect}
\author{
W. F. Kao\thanks{%
gore@mail.nctu.edu.tw} \\
Institute of Physics, Chiao Tung University, Hsinchu, Taiwan}
%\date{June 6, 2003}

\begin{abstract}
Modified Newtonian dynamics can be considered as an effect derived
from a squeezable extra dimension space. The third law of
Newtonian dynamics can be managed to remain valid in the 5-space.
The critical acceleration parameter $a_0$ appears naturally as the
bulk acceleration that has to do with the expanding universe in
this setup. A simple toy model is presented in this Letter to show
that consistent theory can be built with the help of the extra
dimensional space.
\end{abstract}
 \maketitle

%\preprint{hep-th/0603311}
\section{INTRODUCTION}

Modified Newtonian dynamics (MOND) was proposed by Milgrom
\cite{m83a}-\cite{sanders} asserting that gravitational field
requires modifications when the gravitational field strength $N$
is weaker than a critical value $a_0$. This has been shown to be a
good candidate as an alternative to cosmic dark matter (Sanders
2001). The phenomenological foundations for MOND are based on two
observational facts: (1) flat asymptotic rotation curve, (2) the
successful Tully-Fisher \cite{tf77} law, $M \sim V^\alpha$ for the
relation between rotation velocity and luminosity observed in many
spiral galaxies. Here $\alpha$ is close to 4.

It was pointed out \cite{m83a}-\cite{sanders} that there exists a
critical acceleration parameter $a_0 = 1.2 \times 10^{-8} {\rm cm
s}^{-2}$ characterizing the turning point of the effective power
law associated with the gravitational field in MOND. Gravitational
field of the following form was suggested
\begin{equation} \label{gm}
g \cdot \mu ({g \over a_0}) =N
\end{equation}
with a function $\mu$ considered as a modified inertial. Here $N$
is the Newtonian gravitational field produced by certain mass
distribution. Milgrom suggests that
\begin{equation}
\mu (g) = {g  \over  \sqrt{1+g^2}}
\end{equation}
and shows that it provides a best fit with many existing
observation data including the rotational curve of many spiral
galaxies. Recently it was shown that a simpler inertial function
of the following form \cite{fam, fam1}
\begin{equation}
\mu (g) = {g  \over  1+g}
\end{equation}
fits better with RC data of Milky Way and NGC3198.

Evidences accumulated \cite{m83a}-\cite{sanders} show that the
theory of MOND is telling us a very important message. Either the
Newtonian force laws do require modification in the weak field
limit or the theory of MOND may just represent some collective
effect of the cosmic dark matter. In both cases, the theory of
MOND deserves more attention in order to reveal the complete
physics underlying these successful fitting results.

There is, however, a known problem with the momentum conservation
law associated with the theory of MOND. One finds that the
conservation of momentum, a result of the action-reaction
principle, can be resolved with the existence of an extra
dimensional space. One will also be able to show that the proposed
critical acceleration $a_0 \sim H_0 c$ might be closely related to
the existence of an extra dimensional space. Here $H_0$ is the
current Hubble constant. In fact, one will show that a spring-like
extra dimensional  space with squeezable thickness can not only
absorb the missing piece of momentum but also provide a natural
accommodation of the expanding universe in a consistent way.

\section{properties of the extra dimension}

Assuming that there exists an extra dimension space with a
squeezable thickness $Z$ related to the induced field strength $g$
and the inertial mass $m$ of the associated test particle. 5-space
and 4-space will denote the five-dimensional space time and
four-dimensional space time respectively. In addition, 3-space,
the hyper surface of the four dimensional spatial space (or the 4D
bulk space), will denote the conventional three spatial geometry
where Newtonian dynamics holds when $N \gg a_0$. Moreover,
$z$-space will also be denoted as the extra fifth dimensional
space and $b$-space will denote the 4D bulk space without its
3-space hyper surface.

For simplicity, one will write the four dimensional spatial
geometry as a two-dimensional $xz$ strip with the $x$-coordinate
as an abbreviate for the conventional $xyz'$ 3-space. \cite{chkm,
mh} For example, each point on the $x$-line represents a
two-dimensional surface. $z$-membrane will also denote the 4D bulk
space without its 3-space hyper surface.

Three assumptions on the properties of the space will be proposed
: (1) One has assumed that the $z$-space has a squeezable
thickness $Z(x)$ related to the interaction with the gravitational
field $g$. In the presence of a strong field $N$, it is assumed to
have a standard thickness $Z(x)=z_0$. (2) One will further assume
that the thickness $Z$ also has to do with the existence of
inertial mass $m$ of the test particle. In particular, one will
assume that there is also a $z$-space  mass $M(x)$ at the same
point $x$ associated with the existence of any particle with
3-space mass $m$. One assumes that the $z$-space mass $M$ only
acts in the $b$-space and is also proportional to 3-space mass $m$
in the strong field limit. Without losing any generality, one sets
$M \to m$ in the strong field limit. (3) Finally, one assumes that
the $z$-space mass density $M(x)/Z(x)$ remains constant for all
$x$ throughout the $b$-space.

Here is how the idea works : Once the field strength $N$ goes
weak, when MOND effect is apparent, the thickness of the extra
dimension will be squeezed appreciably in response to the
decreasing of $N$ acting on the 3-space world. As a result, part
of its tensional force (or $M$) in the $b$-space is released to
the 3-space world such that the third law of Newtonian dynamics is
preserved in 5-space.

To be more specific, one will show how this could be working
closely with the real world in a consistent way by studying the
simple model proposed by Famaey and Binney \cite{fam, fam1}. Any
different models can also be shown to give similar results with
some necessary modification in the intermediate field strength
region where the strength of $N$ is close to the critical value
$a_0$. Indeed, one can show that the MOND field strength $g$ is
\begin{equation}
g={ \sqrt{N^2+4a_0N} +N \over 2}
\end{equation}
when the Newtonian field strength produced by a source $m_0$ with
a 3-space mass $m_0$. Let us assume that $m_0$ is a point mass
such that $N=Gm_0/r^2$ for simplicity for the moment.
Generalization to any mass distribution is straightforward. The
force $F$ acting on the particle $m$ at a distance $r$ to $m_0$ is
$F=mg$. On the other hand, the test particle $m$ will also produce
a Newtonian field strength $N_0$ at the location of the particle
$m_0$. It is known that $m$ will produce a force $F_0=m_0g(N \to
N_0)$ on $m_0$. Note that one has $N=Gm_0/r^2$ and $N_0=Gm_1/r^2$
acting in the 3-space world. It is apparent that the Newton's
third law of dynamics $m_0 {\bf g}_0 +m {\bf g}=0$ fails to be
observed if the theory of MOND is the whole story. Here $g$ and
$g_0$ represent the MOND field strength acting at $m$ and $m_0$
respectively.

There have been many stimulating research activities
\cite{mh}-\cite{kao051} trying to find a covariant field theory
that is capable of resolving this problem. Instead of these
approaches, we will try a simpler approach. Indeed, one possible
resolution to this dilemma is to assume that the third law of the
Newtonian dynamics is still obeyed in the 5-dimensional space by
assuming that part of the 3-space force $F$ goes into the extra
$z$-dimension. This can be done by assuming
\begin{equation}
{\bf F}_5 = {\bf F}_z + {\bf F}= m{\bf N}
\end{equation}
throughout the 5-space. If this is true, the total 5-dimensional
force ${\bf F}_5=m{\bf N}$ will automatically follow the Newton's
third law. Assume that the $z$-force ${\bf F}_z$ is a repelling
force directing outward from the source $m_0$ and ${\bf F}$ is
pointing inward representing an attracting force. Note that these
directions are all perpendicular to the normal direction of the
3-space hyper surface, i.e.  outward and inward directions do not
have any projection in the $z$-direction. One can show that the
magnitude of ${\bf F}_z$ is that
\begin{equation}
F_z= m { \sqrt{N^2+4a_0N} -N  \over 2}.
\end{equation}
Note that we have assumed that : 1) $F$ only acts on the 3-space
world responsible for the dynamical motion of the particle with
inertial mass $m$ moving in the 3-space world. 2) Similarly, the
force $F_z$ only acts on the $b$-space with inertial mass $M$
responsible for the expansion or contraction of the $xz$-membrane.
$F_z$ is supposed to do nothing directly on the particle's
inertial mass $m$ on the 3-space world.
%even the union of
%$b$-space and 3-space is the complete 4D bulk space.

Writing $F=mg_z$, with $g_z$ as a parameter of acceleration, one
can show that
\begin{equation}
g_z = { \sqrt{N^2+4a_0N} -N  \over 2} .
\end{equation}
It is straightforward to show that
\begin{eqnarray}
g_z  & \to & a_0 , \,\,{\rm if} \, N \gg a_0  \nonumber \\
     & \to & \sqrt{Na_0}, \,\, {\rm if } \, N \ll a_0
\end{eqnarray}
Therefore, one finds that the critical parameter $a_0$ shows up
naturally as a physical parameter of the bulk acceleration acting
on the $xz$-membrane.

Assumption (4): In addition to the first three properties of the
bulk geometry assumed earlier in this Letter, one will also assume
that the bulk $xz$-space is acting as a spring-like membrane with
a small thickness $Z(x)=z_0$ when the field strength $N \gg a_0$.
The thickness $Z(x)$ of the the $z$-space will be squeezed in
response to the change of the field strength $g$ such that
$F_z=Ma_0$ is observed for all $x$ in the 3-space world. Note that
one has assumed that $M(x)$ is the total bulk mass of the membrane
segment associated with the presence of the point mass $m$ at $x$.
One has also assumed that $M(x)=mZ(x)/z_0$ is proportional to the
thickness of the the $z$-space such that $M(x)=m$, up to some
proportional constant set as 1, in the strong $N$ limit.
Therefore, one can show that
\begin{equation}
Z(x)= { \sqrt{N^2+4a_0N} -N  \over 2a_0} z_0 .
\end{equation}
such that $F_z= Ma_0$ is strictly obeyed. In another word, the
whole $b$-space membrane is experiencing a constance acceleration
$a_0$ everywhere outward from the source $m_0$ at $x_0$.
Similarly, point mass $m_0$ will also see a constant acceleration
$a_0$ outward from the source $m$ at $x$. Precisely, each particle
of the system sees a $b$-space acceleration $a_0$ repelling each
other.

In Fig. 1, thin line represents the function $g(N)$ with $g$ and
$N$ both written in unit of $a_0$. In addition, thick line
represents the function $Z(N)$ with $Z$ and $N$ written in unit of
$z_0$ and $a_0$ respectively. In fact, the thick line
$Z(N)=g(N)-N$ in this set of unit.

\begin{figure*}
\includegraphics{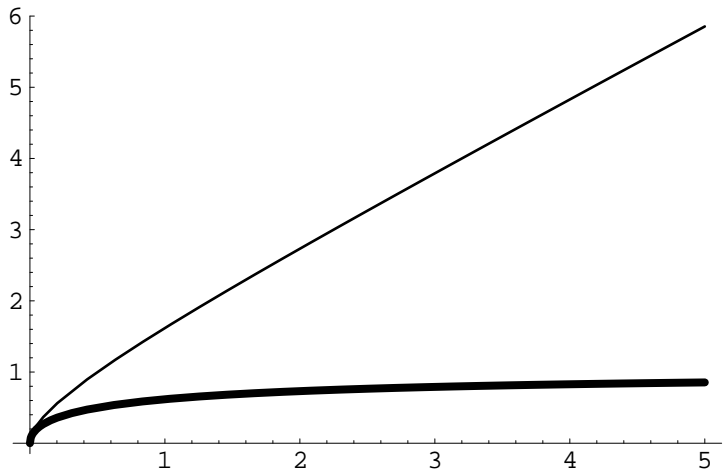}% Here is how to import EPS art
\caption{\label{fig:epsart} Thin line represents the function
$g(N)$ with $g$ and $N$ both written in unit of $a_0$. Thick line
represents the function $Z(N)$ with $Z$ and $N$ written in unit of
$z_0$ and $a_0$ respectively. Note that $Z \to 1$ as $N \gg 1$ in
this unit.}
\end{figure*}

One has to assume that the bulk-space acts like a spring pulling
against the repelling bulk forces and holds the space time from
accelerating far apart. The spring tension acts in response to the
repelling force in a consistent way such that the total force
acting on the $z$-membrane is cancelled or close to nothing. The
resulting universe is therefore seen to only undergo a constant
expansion or a mild accelerating expansion accordingly. This
situation is rather similar to an incompressible pressurized fluid
in a closed container that has constant pressure everywhere.

Since the accelerating parameter $a_0$ associated with the
$z$-space has to do with the expansion of the universe, it should
be closely related to the unique physical parameter of the
expanding universe: the Hubble constant $H_0$. Therefore, it is
nature to expect $a_0 \sim H_0 c$ based on dimensional argument.
Even one does not know how to relate this relation formally in a
rigorous way. This question may also be related to the question we
did not answer in this Letter: why the thickness $Z(x)$ of the
$z$-space takes the form given by Eq. (9) or any other form
prescribed by a different alternative model. Our result only shows
that the law of momentum conservation can be secure in a
consistent way with the help of the proposed fifth dimensional
bulk space with finite thickness. Hopefully, in light of the
consistent setup shown in this Letter, one will soon be able to
find a more rigorous way to derive these relations from a formal
field theoretical approach in the near future.

\section{Conclusion}

One has made totally four assumptions on the properties of the
bulk $z$-space: (1) it has a finite thickness $Z(x)$, (2) it has
an inertial mass $M(x)$ in response to the force $F_z$ acting on
the bulk space, (3) bulk mass density $M/Z=$ constant throughout
the bulk space, (4) the force acting on the $b$-space $F_z=Ma_0$
tends to be acting on an "incompressible" fluid such that particle
far apart tends to accelerate outward against each other with a
constant acceleration $a_0$.

As a result, one derives the dependence of $Z(x)$ shown in Eq.
(9). It indicates that part of the effect of the mass $M$ in the
$b$-space is released to the 3-space world in compensation as a
deformation of the inertial $\mu(g/a_0)$ acting on the 3-space
world. All forces adding together are therefore managed to obey
the third law of Newtonian dynamics.

The first three assumptions are quite reasonable assumptions one
can imagine for a well behaved membrane. The forth assumption
asserts that the membrane tends to distribute forces acting on it
equally such that constant acceleration is achieved everywhere on
the bulk space. For example, two distinct galaxies tend to expand
or accelerate outward from each other. This property is similar to
the equal pressure acting on an ideal fluid enclosed in a closed
container. Therefore, the forth assumption is also a reasonable
one.

The result shown here with a simple toy model indicates that the
theory of MOND might have to do with the input from the extra
dimensional space. The expansion of the bulk space (along the
3-space direction $x$) is achieved as a bonus of the toy model.
Assumptions made on the properties of the extra dimensional space
are considered to be reasonable ones.

Therefore, it appears that the somewhat successful theory of MOND
does have close relation to the effect from the extra dimension.
One cannot rule out the possibility that the theory of MOND is
merely a collective effect of the cosmic dark matter working
closely following the models proposed by Milgrom, Famaey \&
Binney, and many others. In any case, the theory of MOND deserves
more attention for a more clear physical picture of the galaxy
dynamics.

{\bf \large Acknowledgments} This work is supported in part by the
National Science Council of Taiwan.

\vspace{ 2cm}

\end{document}